\documentclass[12pt]{article}
\pdfoutput=1
\usepackage{graphicx,rotating,hyperref,slashed,amsmath,xcolor,amssymb,amsfonts,colortbl,cite}
\makeatletter
\graphicspath{{plots/}}%here the directory where it looks for the figures

\hypersetup{colorlinks,bookmarksopen,bookmarksnumbered,
linkcolor=blus,pdfstartview=FitH,urlcolor=rossos,citecolor=verde}
\allowdisplaybreaks		
\newcommand{\plu}{\scalebox{0.9}[1.0]{+}}

\def\beq{\begin{equation}}
\def\eeq{\end{equation}}

\def\be{\begin{equation}}
\def\ee{\end{equation}}

\def\bea{\begin{eqnarray}}
\def\eea{\end{eqnarray}}

\def\0{{\boldsymbol 0}}

\def\lsim{\mathrel{\rlap{\lower3pt\hbox{\hskip0pt$\sim$}}
   \raise1pt\hbox{$<$}}}         %less than or approx. symbol
\def\gsim{\mathrel{\rlap{\lower4pt\hbox{\hskip1pt$\sim$}}
   \raise1pt\hbox{$>$}}}         %greater than or approx. symbol

 \newcommand{\sfootnote}[1]{}
\definecolor{bluc}{cmyk}{1,1,0,0.1}
\definecolor{rossoCP3}{cmyk}{0,.88,.77,.40}
\definecolor{rosso}{cmyk}{0,1,1,0.4}
\definecolor{rossos}{cmyk}{0,1,1,0.55}
\definecolor{rossoc}{cmyk}{0,1,1,0.2}
\definecolor{verdes}{cmyk}{0.92,0,0.59,0.4}

\hypersetup{colorlinks, bookmarksopen, bookmarksnumbered,
citecolor=verdes, linkcolor=bluc, pdfstartview=FitH, urlcolor=rossos}

\newcommand{\mio}[1]{}

\definecolor{Gray}{gray}{0.95}

\usepackage{multicol}
\usepackage{color}
\definecolor{rosso}{cmyk}{0,1,1,0.4}
\definecolor{rossos}{cmyk}{0,1,1,0.55}
\definecolor{rossoc}{cmyk}{0,1,1,0.2}
\definecolor{blu}{cmyk}{1,1,0,0.3}
\definecolor{blus}{cmyk}{1,1,0,0.6}
\definecolor{bluc}{cmyk}{1,1,0,0.1}
\definecolor{verde}{cmyk}{0.92,0,0.59,0.25}
\definecolor{verdec}{cmyk}{0.92,0,0.59,0.15}
\definecolor{verdes}{cmyk}{0.92,0,0.59,0.4}

\def\circa#1{\,\raise.3ex\hbox{$#1$\kern-.75em\lower1ex\hbox{$\sim$}}\,}

\newfam\rsfsfam
\def\mathscr#1{{\fam\rsfsfam\relax#1}}

\def\circa#1{\,\raise.3ex\hbox{$#1$\kern-.75em\lower1ex\hbox{$\sim$}}\,}
\makeatletter

\def\hhref#1{\href{http://arxiv.org/abs/#1}{arXiv:#1}} % in bibliography

\newcommand{\doi}[1]{\href{http://dx.doi.org/#1}{[doi]}}

\def\hhref#1{\href{http://arxiv.org/abs/#1}{arXiv:#1}} 
 
\def\art{\@ifnextchar[{\eart}{\oart}}
\def\eart[#1]#2#3#4#5#6{{\rm #2}, {\em #3 \bf #4} {\rm (#6) #5} ({\em #1})}

\def\article{\@ifnextchar[{\earticle}{\oarticle}}
\def\oarticle#1#2#3#4#5#6{{\rm #1}, {\em ``#6''}, {\rm #2 #3 (#5) #4}}
\def\earticle[#1]#2#3#4#5#6#7{{\rm #2}, {\em ``#7''}, {\rm #3 #4 (#6) #5}  [\hhref{#1}]}
\def\hepart[#1]#2{{\rm #2, \em#1}}
\def\heparticle[#1]#2#3{#2, {\em ``#3''} [\hhref{#1}]}

\newcounter{alphaequation}[equation]
\def\thealphaequation{\theequation\hbox to
0.6em{\hfil\alph{alphaequation}\hfil}}
\def\eqnsystem#1{
\def\@eqnnum{{\rm (\thealphaequation)}}
\def\@@eqncr{\let\@tempa\relax \ifcase\@eqcnt \def\@tempa{& & &} \or
  \def\@tempa{& &}\or \def\@tempa{&}\fi\@tempa
  \if@eqnsw\@eqnnum\refstepcounter{alphaequation}\fi
\global\@eqnswtrue\global\@eqcnt=0\cr}
\refstepcounter{equation} \let\@currentlabel\theequation \def\@tempb{#1}
\ifx\@tempb\empty\else\label{#1}\fi
\refstepcounter{alphaequation}
\let\@currentlabel\thealphaequation
\global\@eqnswtrue\global\@eqcnt=0 \tabskip\@centering\let\\=\@eqncr
$$\halign to \displaywidth\bgroup \@eqnsel\hskip\@centering
$\displaystyle\tabskip\z@{##}$&\global\@eqcnt\@ne
\hskip2\arraycolsep\hfil${##}$\hfil& \global\@eqcnt\tw@\hskip2\arraycolsep
$\displaystyle\tabskip\z@{##}$\hfil
\tabskip\@centering&\llap{##}\tabskip\z@\cr}
\def\endeqnsystem{\@@eqncr\egroup$$\global\@ignoretrue} \makeatother

\definecolor{fiorentina}{rgb}{.5,0,.5}

\begin{document}

\vspace{1truecm}
 
\begin{center}
\boldmath

{\textbf{\Large On long range 
axion hairs for  black holes}}

\unboldmath

\unboldmath

\bigskip\bigskip

%\vspace{0.4truecm}
\vspace{0.1truecm}

{\bf Francesco Filippini, Gianmassimo Tasinato}
 \\[8mm]
{\it  Department of Physics, Swansea University, Swansea, SA2 8PP, UK}\\[1mm]

\date{today}

\vspace{1cm}

\thispagestyle{empty}
{\large\bf\color{blus} Abstract}
\begin{quote}
The physics of black holes can suggest new ways to test the existence of  axions. 
 Much work has been done so far to analyse
the phenomenon of superradiance  associated with   axions in the ergoregion surrounding rotating black holes. In this work,   we  instead investigate how
Chern-Simons axion couplings of the form $\phi \,F\,\tilde F$ and $\phi \,R\,\tilde R$, 
well motivated by  particle physics and string theory,      can induce long range profiles for light
axion fields 
 around charged  black holes, with or without spin.  We extend known  solutions describing axion hairs around  spherically  symmetric, asymptotically flat dyonic  black hole configurations,
  charged under    $U(1)$ gauge symmetries, by  including non-minimal couplings
  with gravity. 
 The axion  acquires  a    profile controlled
 by the black hole conserved charges,  and
we analytically determine how it 
 influences the black hole horizon and its properties. We find 
a Smarr formula  applying to our configurations. 
  We then    generalise  known solutions describing axion hairs around slowly rotating  black hole configurations
   with  charge. To make contact with phenomenology, we briefly study  how
   long range  axion profiles induce  
    polarised deflection of light rays, and the properties of
   ISCOs for the black hole configurations we investigate.

\end{quote}
\thispagestyle{empty}
\end{center}

\setcounter{page}{1}
\setcounter{footnote}{0}

%\newpage

%\tableofcontents

\bigskip

\section{Introduction}

Axions can be considered among the most well motivated candidates for physics   beyond the Standard Model. Introduced
in order to solve the strong CP problem \cite{Peccei:1977hh,Peccei:1977ur,Weinberg:1977ma,Wilczek:1977pj} (see 
the review \cite{Kim:2008hd}), it was soon
realised they have relevant implications for cosmology, as
dark matter  \cite{Abbott:1982af,Dine:1982ah,Preskill:1982cy} or dark energy 
candidates \cite{Carroll:1998zi}  (see the review  \cite{Marsh:2015xka}). Moreover, axion fields
arise naturally in string theory constructions (see e.g.  \cite{Svrcek:2006yi}) with a broad
variety of couplings and masses,  motivating
the string axiverse scenario of \cite{Arvanitaki:2009fg}. The quest 
of their  experimental detection is currently an active
research field, see e.g.
 \cite{Hook:2018dlk} for a recent pedagogical review. 
Very light axion fields are  specially  interesting dark matter candidates  \cite{Turner:1983he,Press:1989id,Sin:1992bg,Hu:2000ke,Goodman:2000tg,Peebles:2000yy,Amendola:2005ad,Schive:2014dra,Hui:2016ltb}, and their specific properties can lead to distinctive observational consequences. 

In this work we investigate 
the relation between the physics of  axions and black holes.
Light axions, whose Compton wavelength is of the order of a black hole Schwarzschild radius,
  can  deposit on the ergosphere of a rotating black hole, and cause instabilities associated with the phenomenon of 
  black hole superradiance \cite{Penrose:1969pc,Press:1972zz,Cardoso:2004nk}. Such instabilities can have observable  implications, 
  since axions can extract rotational
 energy from spinning black holes  \cite{Detweiler:1980uk,Cardoso:2005vk,Dolan:2007mj,Arvanitaki:2010sy} (see \cite{Brito:2015oca} for a comprehensive
 review), and  lead to a distinctive emission of detectable gravitational waves  (see e.g. \cite{Arvanitaki:2014wva,Brito:2017wnc}) \footnote{
  Possible ways to probe properties
of axions with neutron stars are also being investigated, see e.g.  \cite{Hook:2017psm,Huang:2018pbu,Garbrecht:2018akc}.
 }. Here we focus
 on the situation where the axion Compton wavelength is {\it much larger} than a black hole Schwarzschild radius. In this regime, the axion $\phi$ can be considered
 effectively massless, and the axion Lagrangian  enjoys a shift symmetry $\phi\to\phi+{\rm const}$. It is then  natural to ask whether $\phi$ can develop an extended   radial profile,  leading to  long range axion hairs around a black hole. There are several no-hair arguments to overcome, starting from Bekenstein results \cite{Bekenstein:1971hc,Bekenstein:1995un}. On the other hand, a key property of  axion fields we can exploit is  
 the fact that they are characterised by Chern-Simons couplings to gauge fields and gravity,
 \be\label{eq_intro}
 \phi\,F \,\tilde F\hskip0.5cm,\hskip0.5cm  \phi\,R \,\tilde R
 \ee
 with $\tilde F$ and $\tilde R$ respectively the Hodge dual of the gauge field strength and Riemann tensor.  While the gauge Chern-Simons coupling
 naturally appears when dealing with anomalous symmetries of the QCD axion (see the reviews \cite{Kim:2008hd,Hook:2018dlk}), the gravitational Chern-Simons coupling
 arises in explicit calculations in  string theory constructions, see e.g. \cite{Svrcek:2006yi}. 
  The linear Chern-Simons axion couplings in eq \eqref{eq_intro} source a non-vanishing 
  asymptotic value for the axion fields for charged and/or rotating black holes, and  allow one to avoid black hole no-hair theorems, somewhat analogously 
  to what happens in Horndenski theories, where linear couplings with the Gauss-Bonnet invariant were found in \cite{Sotiriou:2013qea,Sotiriou:2014pfa} to be an important
  ingredient for overcoming the no-hair
  theorem of \cite{Hui:2012qt}. 
  Our work proceeds as follows:
  \begin{itemize}
  \item In Section \ref{sec_cons} we set the stage  presenting   the shift symmetric axion system we consider. We include additional  derivative couplings 
of the axion field with  curvature, allowed by the symmetries of the system,   which  can be motivated by high energy constructions. We explain in detail why the Chern-Simons
couplings \eqref{eq_intro} are expected to induce   regular axion hairs around  regular spherically symmetric charged black holes, and around  rotating
black holes (with or without charge). 
\item In  Section \ref{sec_spher} we  study spherically  symmetric dyonic black holes, with electric and magnetic charges, equipped by a long
range axion hair. We generalise known solutions \cite{Campbell:1991rz,Campbell:1991kz,Lee:1991jw} to  cases  where the axion
 has derivative couplings with the Ricci tensor, motivated in Section \ref{sec_cons}, showing that Chern-Simons couplings can be  effective
 in generating axion hairs also in theories with   non-minimal derivative couplings of axions to curvature.  
 The axion profile has a `secondary' charge,   i.e. it depends on the black hole conserved charges.  
 We work at leading order 
 in an expansion of the relevant coupling constants, finding analytical solutions and discussing  how the properties of the black hole horizon are
 modified by the axion hair.
 \item In Section  \ref{sec_rot} we consider slowly rotating charged black holes, finding new configurations with axion hairs, which generalize
 and interpolate between known solutions \cite{Campbell:1990ai,Boskovic:2018lkj}. Interestingly, we find that the gauge and gravitational Chern-Simons couplings symultaneously 
 play a role in characterising the axion hair and the geometry, leading to mixed contributions depending on both these parameters.  
  \item In Section  \ref{sec_app} we study phenomenological applications of our findings, showing how long range black hole
  axion hairs can bend differently the two polarizations of light crossing in proximity of the black hole (elaborating on the methods of \cite{Plascencia:2017kca}), and
  by studying properties of the innermost stable circular orbits  (ISCOs)  for our configurations.
  \end{itemize}

\section{System under consideration}\label{sec_cons}

\subsection{The action}

We consider the following action describing gravity coupled with a $U(1)$ gauge field $A_\mu$ (with $F_{\mu\nu }
\,=\,\partial_\mu A_\nu-\partial_\nu A_\mu$ its field strength) and a pseudoscalar
axion 
field 
$\phi$ 
%(we set $M_{Pl}^2\,=\,1/2$): 
%\begin{itemize}
%\item  The action we consider is 
\be
S\,=\,\int d^4x \,\sqrt{-g}\,\left[ \frac{R}{4}-\frac{1}{4}F^{\mu\nu}F_{\mu\nu}-
{\cal L}_{\phi} \left(\phi,\partial \phi \right)
%\frac{1}{2}\nabla^{\alpha}\phi\nabla_{\alpha}\phi -\frac{\mu^2}{2}\phi^2
 -\frac{g_F}{4}\phi F_{\mu\nu} \tilde{F}^{\mu\nu}-\frac{g_R}{4}\phi {R}_{\mu\nu\rho\sigma} \tilde R^{\mu\nu\rho\sigma} \right] \,,\label{eq:ACTION}
\ee
 where from now on we have set units such that  $c\,=\,G_N\,=\,1$.  
This action describes Einstein gravity coupled with a $U(1)$ Maxwell field, and a pseudoscalar axion described by the Lagrangian density ${\cal L}_{\phi}$
 (to be specified in what follows).
 The  axion 
is also non-minimally coupled with gauge and metric fields through Chern-Simons operators weighted by the constants $g_F$ and $g_R$: the dual of the 
field strength and Riemann tensor are defined as $\tilde F^{\mu\nu}\,=\frac{1}{2}\,\,\epsilon^{\mu\nu\rho\sigma}\,F_{\rho \sigma}$ and $\tilde R^{\mu\nu \rho \sigma}\,=\frac{1}{2}\,\epsilon^{\mu \nu\alpha \beta} R^{\rho\sigma}_{\,\,\,\;\;\;\alpha\beta} $. 
The axion can be identified with the Nambu-Goldstone boson of a spontaneously
broken Peccei-Quinn symmetry, or with the Hodge dual of a three-form which appear
by string theory scenarios. In both cases, Chern-Simons couplings with
 the structure  above arise naturally.   The  dimensionful     coupling constants 
 of the Chern-Simons terms 
 can be associated
 with the axion decay constants: for our purposes we 
 appropriately weight those operators with powers of some fundamental mass scale (for example the Planck mass)  and  
 consider the quantities $g_{F,\,R}$ as dimensionless free constant   parameters, 
 influencing the geometry of the space-times we  consider.

At the perturbative level, axion fields typically enjoy a shift symmetry $\phi\to\phi\,+\,{\rm const}$,  which can be broken by non-perturbative effects giving a mass to the axion.  For cosmological purposes
 --  as dark energy or dark matter candidates -- axions are typically very light.
  For the aim  of this work, we   consider axions whose Compton wavelength is much larger
than the black hole size. Being the Compton wavelength of an axion field of mass $m_\phi$
\be
\lambda_{\rm Compton}\simeq \,\left(\frac{10^{-10}\,{\rm eV}}{m_\phi} \right)\,\left( \frac{M_{BH}}{M_{\odot}}\right)\, r_{\rm Schw}^{\rm sun}\,,
\ee
%{\bf \color{red} check carefully the previous formula: i got $r_{compton}$ from Urbanos paper, while
%solar schwarschild radius from wikipedia. }
we  can assume that the axion masses are  much smaller than  $10^{-10}$ eV when considering solar mass
black holes, or much smaller $10^{-16}$ eV when investigating supermassive  black holes sitting in the centre of galaxies,
so that $\lambda_{\rm Compton}$ is well larger than the corresponding black hole Schwarzschild radius. 
%Hence, for our purposes, 
% {\bf \color{red} put  in the intro the formula for the compton} 
 As a consequence, we will  neglect shift symmetry-breaking
 effects, and   
 consider the axion field  as effectively massless. 
  The shift-symmetric axion Lagrangian ${\cal L}_\phi$ we shall consider  is quadratic in the axion field, so to respect the pseudoscalar
  properties of the axion. It  contains the standard axion kinetic terms, plus non-minimal
 derivative couplings with gravity:
 \be \label{scal_lag}
 {\cal L}_{\phi}\,=\,\frac{1}{2}\nabla^{\mu}\phi\nabla_{\mu}\phi +\lambda\,G^{\mu\nu}\,\nabla_\mu \phi \nabla_\nu \phi, 
 \ee
 with $G^{\mu\nu}$ being the Einstein tensor. This is the most general   shift symmetric
 quadratic action for the pseudoscalar axion field, which is invariant under the parity symmetry properties
 of the pseudoscalar axion, and leads to second order equations of motion.  
 The second contribution in (\ref{scal_lag}) is a dimension-six operator belonging to the Horndeski action. We weight it with appropriate powers of the Planck mass, and consider the
 parameter $\lambda$ as dimensionless. 
 
  We include the  operator controlled by $\lambda$ for the following reasons:
  \begin{itemize}
  \item[i)]
   It is  allowed by the symmetries
 of the  axion system  -- shift symmetry and parity invariance --  hence it can be expected from an effective
 field theory point of view,  being induced by perturbative loop corrections to the tree-level action.
  Although derivative couplings of axions with curvature are rarely considered (see however \cite{Folkerts:2013tua} for
  an example), 
 they can be motivated by string theory constructions. For example, \cite{Metsaev:1987zx} computed
 $\alpha'$ and loop corrections in heterotic string systems including the 3-form antisymmetric tensor field $H_{\mu\nu\rho}$, finding
 various couplings of the latter with Riemann and Ricci tensors. Once the three form field is Hodge-dualised to a pseudoscalar, such couplings can  lead to derivative
 couplings of the axion to curvature tensors~\footnote{These are heuristic, preliminary  considerations that would be interesting
 to further develop  in a proper string theory setting; on the other hand, this goes beyond the
 scope of this work.}, as in Lagrangian \eqref{scal_lag}. 
 \item[ii)]
  Non-minimal couplings of scalar fields with curvature have a long history  in models of dark energy (see 
 e.g. \cite{Clifton:2011jh} for a review)
 and cosmological inflation (as in Starobinsky \cite{Starobinsky:1980te} or Higgs inflation \cite{Bezrukov:2007ep,Germani:2010gm}). Although they are typically
 suppressed by some high energy scale, they can have important implications
 when the scalar acquires a non-trivial $vev$.
 Their consequences
 for black hole physics, and the existence of scalar hairs evading
 strong no-hair theorems,  is a well studied subject. 
 It is then natural to  investigate  this topic  for the case of pseudoscalar axions, and  
  to explore the role of Chern-Simons couplings in allowing for black hole axion hairs,  including
   derivative couplings with gravity.
\end{itemize}
  In the next subsection, we  explain why the gauge and Chern-Simons couplings 
  play an essential role in allowing axion hairs on charged and rotating black holes. 
  
\subsection{Long range axion hairs and black holes}\label{sec_hairs}
 
%  A light axion can acquire a long range profile, and influence spherically symmetric configurations.
   %For example, 
   Cosmological considerations motivate the investigation of  light axion fields
   with Compton wavelengths well larger than the typical sizes of astrophysical black
   holes. In this context,  the axion field
 can be considered as massless; it might acquire a long range profile,  influencing  spherically symmetric configurations. 
  But can a regular, asymptotically flat black hole with extended   axion hairs exist?
   There is a long list of  no-hair theorems to overcome, starting from Bekenstein results \cite{Bekenstein:1971hc,Bekenstein:1995un}, which
   forbid configurations with a long range (pseudo)scalar $\phi$; no-hair arguments   
     are particularly  powerful in the case of shift-symmetric Lagrangians \cite{Hui:2012qt} like ours. (See  e.g. 
    \cite{Herdeiro:2015waa,Sotiriou:2015pka,Volkov:2016ehx} for reviews.) It is then interesting, both theoretically and phenomenologically, to assess
    whether  long range axion hairs 
    exist around black holes. 
      
      The theorem  discussed in \cite{Hui:2012qt}  makes use of the properties of the conserved current $J^\mu$ associated
      with the shift symmetry:  \cite{Hui:2012qt}  
      shows that -- for   systems that do not contain linear terms in $\phi$ -- regularity of the current
      at the horizon imposes a trivial radial profile for $\phi$. On the other hand, 
      in the context of Horndeski theories, a way out has been found in  \cite{Sotiriou:2013qea} by noticing that the 
      quintic Horndeski Lagrangian contains   contributions linear in $\phi$ (coupling the field with 
      the Gauss-Bonnet combination), hence evading the theorem and allowing 
      for black hole hairs in asymptotically flat configurations. (Exact solutions
      in related systems were found in \cite{Kanti:1995vq}.) 
      
      In our shift-symmetric context,  action \eqref{eq:ACTION}, we do as well have linear couplings of the 
      axion field with  gauge and gravitational Chern-Simons terms~\footnote{Shift symmetry is
      ensured by the fact that $F \tilde F$ and $R \tilde R$ are by themselves  total derivatives.}, 
      hence we can in principle find black holes
      with long range axion hairs.    The scope of our work is indeed  to 
       %one of our purposes is to 
       demonstrate that  these couplings offer   new opportunities  to find regular,
       asymptotically flat    black hole solutions with long-range
         pseudoscalar fields.    
      Examples of black hole solutions with axion hairs, with and without
      charge and spin, have indeed been determined  
      in the past \cite{Campbell:1990ai,Campbell:1991rz,Campbell:1991kz,Lee:1991jw,Reuter:1991cb}, and the subject has been particularly developed in the context
      of Chern-Simons gravity \cite{Jackiw:2003pm} (see e.g. \cite{Yunes:2009hc} and
      the review \cite{Alexander:2009tp}). 
      
        In our work, we further generalise
       these       solutions     at the light of these considerations,   and extend them to the case of further derivative couplings of the axion with gravity,
       described by Lagrangian \eqref{scal_lag}, and by 
       considering the simultaneous presence of $g_F$, $g_R$.  
      The presence of  Chern-Simons terms leads to sources for the axion field which are
 %Chern-Simons contributions are 
 non-vanishing  at  spatial infinity,  inevitably inducing  long-range axion profiles around black holes.
   In particular, the $F \tilde F$ coupling provides a source 
 for spherically symmetric dyonic black
holes, with  electric ($Q$) and magnetic ($P$) charges, since it  
%this operator is non-vanishing at infinity, and 
acquires the
following profile asymptotically far from the black hole:
\be \label{souF}
F \tilde F\,\sim\,\frac{P\,Q}{r^4}\,.
\ee
  On the other hand, the gravitational $R \tilde R$ coupling sources an axion radial profile  when the black hole 
geometry is axisymmetric, for example when it is rotating 
with rotation parameter $a=J/M$, ($J$  being the angular momentum  and $M$ the black
hole mass). In the 
 uncharged case, and
in the small rotation limit ($a\ll1$),  one finds 
 \be\label{souR}
 R \tilde R\,\sim\,\frac{a\,M^2}{r^7}\cos \theta\,.
 \ee
 The resulting axion field induced by the sources 
   \eqref{souF} and \eqref{souR}  backreacts on the geometry 
 and influences the physical properties of the black hole at the position of the horizon.

  In the next Sections, we first study charged spherically symmetric black hole configurations for this system; then
  we study slowly rotating charged solutions, and afterwards we analyse  phenomenological
  consequences of our findings.

\section{Spherically symmetric configurations}\label{sec_spher}

In this Section, we discuss  asymptotically flat, spherically symmetric configurations with axion hairs around
 dyonic black holes, characterised by both magnetic and electric charges associated with the $U(1)$
 gauge field $A_\mu$.  This gauge field can be associated with some
  gauge symmetry beyond the Standard Model -- for example associated with dark gauge bosons for dark
  matter interactions  \cite{Ackerman:mha,Feng:2009mn,Agrawal:2016quu}, or some additional gauge
  group motivated by string theory -- 
   or with standard electromagnetism: our
  discussion applies to both cases \footnote{The configurations we consider differ from string theory configurations found in 
   dilaton--Maxwell-gravity systems and 
  dilaton-axion-Maxwell-gravity systems (see e.g. respectively \cite{Gibbons:1987ps}, \cite{Shapere:1991ta}
   and
  the comprehensive review \cite{Ortin:2015hya}), since  in our case we do not have any scalar dilaton in our system, but only
  a pseudoscalar axion. }. 
 The Chern-Simons gauge
 coupling $F \tilde F$ sources an axion profile that scales as  $1/r$ 
 asymptotically far from the object (while the gravitational
 Chern-Simons term $R\tilde R$ vanishes for spherically symmetric
 solutions). Its  backreaction on the geometry
 and the gauge field can be analytically computed in a perturbative
 expansion   in the dimensionless coupling constants 
 $g_F$ and $\lambda$, appearing in action \eqref{eq:ACTION} and scalar Lagrangian
 \eqref{scal_lag}.  We work at leading order in a perturbative expansion on these 
 parameters to maintain our expressions relatively simple, and since interesting physical
 phenomena already appear 
at leading order in such expansion~\footnote{For completeness, in Appendix \ref{app_a} we present an
exact solution for a specific, potentially large value of the coupling constant $\lambda$.}.

 The axion profile  affects the electric field,
 but not the magnetic one.  Our solution generalises the spherically symmetric configurations 
 studied by Campbell et al  \cite{Campbell:1991rz,Campbell:1991kz} and by Lee and Weinberg \cite{Lee:1991jw} (see also
 \cite{Reuter:1991cb}) by including non-minimal couplings between axion and metric, described
 by the Horndeski coupling $\lambda$ in the Lagrangian density \eqref{scal_lag}. As we shall discuss, the properties
  of our configurations are  different from other black hole solutions in
  theories with non-minimal couplings with gravity. 
   %We will
 %present our solutions  at leading
 %order in an expansion of the equations of motion in the
 %dimensionless  parameters  $g_F$ and in $\lambda$.
 % already
 %in this approximation the effects of the new couplings become apparent. 

 \bigskip
 \noindent  {\underline{\bf The axion profile}} 
 At leading order in an expansion in both $g_F$ and $\lambda$, 
  the unique solution for the pseudoscalar equation of motion
 that is regular at the black hole horizon is
 \beq 
\phi(r)\,=\,-g_F\,P\, Q \left[\frac{\log
   \left(1-\frac{r_{-}}{r}\right)}{r_{-}
   r_{+}}-\frac{\lambda}{r_{-}^4} \left(\frac{2\,r_{-}}{r
   }+\frac{r_{-}^2}{r^2 }+\frac{2\,r_{-}^3}{3
   r^3 }+\frac{r_{-}^4}{2 r^4}+2 \log
   \left(1-\frac{r_{-}}{r}\right)\right)\right]\,, \label{eq:scHAI}
\eeq
where $r_{\pm}$ correspond to the outer/inner horizons of Reissner-N\"ordstrom  (RN) black
holes of mass $M$ with electric and magnetic charges, respectively $Q$ and $P$ (recall that 
we are choosing units so that $M_{\rm Pl}^2\,=\,1/2$):
\be \label{old_hor}
r_{\pm}=M\pm\sqrt{M^2-P^2-Q^2}\,.
\ee 
In writing the  solution  \eqref{eq:scHAI}, we fix integration constants such to impose regularity
of this configuration at asymptotic infinity, and at the external RN horizon $r_+$;
  we could add  to this expression a constant $\phi_0$ which however has no physical consequences
due to shift-symmetry. 
The  axion profile corresponds to   a `secondary' hair, since there is no independent pseudoscalar charge but it 
is fully controlled by the black hole charges ($M,\,P,\,Q$); on the other
hand, as we shall see, the axion affects properties
of the black hole horizon.   Notice that the axion profile is regular
at the position $r_+$ of the outer horizon, while it diverges at the inner horizon $r_-$ (more on this later).  
Expanding for large values of $r$, one finds~\footnote{It is straightforward  to verify
 that this asymptotic, large $r$ profile for the scalar field remains valid also beyond a leading
order expansion in the coupling constants $g_F$, $\lambda$. We will use this fact in Section \ref{sec_pheno_1}. 
 } 
\be \label{scal_asympt}
\phi\sim \left(\frac{g_F\,P\, Q}{r_+}\right)\,\frac{1}{r}+{\cal O}\left(\frac{1}{r^2} \right)\,.
\ee
Interestingly, we find that the pseudoscalar axion profile scales as $1/r$, so charged black holes
with  gauge couplings $ \phi F \tilde F$ activate a monopolar axion configurations. This is a difference
with respect to
% contrarily toqualitative
 what happens with gravitational Chern-Simons couplings  $ \phi R \tilde R$, as we shall
 learn in Section \ref{sec_rot},  where the axion radial profile
 scales at least as $1/r^2$ \cite{Campbell:1990ai}. This  axion profile has interesting consequences
 for the physics of the horizon and for  the phenomenology of the system. 
 Notice that although the axion scales with the radial coordinate as $1/r$, it can more
 easily circumvent fifth force constraints   \cite{Carroll:1998zi,Adelberger:2003zx}, since being a pseudoscalar
 it does not directly couple to the trace of matter energy momentum tensor. Moreover, one
 can check that the corrections weighted by $\lambda$, associated with derivative couplings with the metric, influence
 the axion profile only starting at order $1/r^4$: this is probably a manifestation of  Vainshtein screening
 in this context (see e.g. \cite{Babichev:2013usa} for a review).

  \bigskip
 \noindent  {\underline{\bf Gauge field solution}} The vector potential for this configuration has both electric and 
 magnetic components 
 %switched on (again, at leading order in a $g_F$, $\lambda$ expansion):
 \be
 A_{\mu}=\left\{ A_t(r),\,0,\,0,\,A_\varphi(\theta) \right\}
 \ee
 where, at leading order in a $g_F$ and $\lambda$ expansion, we find the solution
 
 \beq  \label{Aprof}\small{
%\hspace*{-1cm}%
\begin{aligned}
A_t(r)=&-\frac{Q}{r}+\frac{g_F^2\,P^2 Q}{r} \biggl[\frac{r_{-} \left(r_{-} r_{+}-2 Q^2\right)+ \left[Q^2 (r_{-}-2 r)+r_{+} r_{-}(r-r_{-})
   \right]\log
   \left(1-\frac{r_{-}}{r}\right)}{r_{-}^3
   r_{+}^2}+\lambda \,a_{\lambda}(r) \biggr]\\[10pt]
 a_{\lambda}(r)=&\frac{1}{30 r^5 r_{-}^2 }  \biggl\{r_{-} r_{+}
   \biggl[60 r^4 (r_{-}+4 r_{+})+30 r^3 r_{-} (r_{-}-4 r_{+})+20 r^2
   r_{-}^2 (r_{-}-2 r_{+})+5 r r_{-}^3 (3 r_{-}-4 r_{+})\\
  & -12
   r_{-}^4 r_{+}\biggr]-5 P^2 \biggl(12 r^4 (r_{-}+5 r_{+})+6 r^3 r_{-}
   (r_{-}-5 r_{+})+2 r^2 r_{-}^2 (2 r_{-}-5 r_{+})+r r_{-}^3 (3
   r_{-}-5 r_{+})\\
   &-3 r_{-}^4 r_{+}\biggr)-60 \frac{r^4}{r_{-}} \log
   \left(1-\frac{r_{-}}{r}\right) \Biggl[Q^2 \Biggl( -r r_{-}-5 r_{+}r+5 r_{-} r_{+} \Biggr)+r_{-} r_{+} (r r_{+}-r_{-} r_{+})\Biggr]\biggr\}\\[10pt]
A_\varphi(\theta)=&\,-P\,\cos \theta. \\
\end{aligned} }
\eeq
 This  corresponds to a magnetic monopole  configuration with magnetic charge $P$, additionally charged
 under an electric field  $E_r\,=\,\partial_r\,A_t(r)$ whose
 profile is modulated by the axion configuration, and by the presence of the  magnetic
 charge $P$. This
  fact
  resembles, in curved space-times, a phenomenon called  Witten effect \cite{Witten:1979ey,Wilczek:1987mv}, where an axion configuration switches on an electric field in a magnetic monopole background  in the presence  of a Chern-Simons coupling $\phi F \tilde F$.

 \bigskip
 \noindent  {\underline{\bf Metric  solution}} 
 %We find that the backreaction of the axion profile on the metric
 %configuration starts at quadratic order in the coupling constant $g_F$. 
At leading order
 in a $g_F$ and $\lambda$ expansion,  the metric line element reads as follows:
 \beq
ds^2=-F(r)\,dt^2\,+\frac{dr^2}{h(r) \,F(r)}+ r^2\,d\theta^2+ r^2\,\sin^2\theta\,d\varphi^2\,,
\eeq
with

\vspace{5mm}
\small{ \beq  \label{met_sol1}
\hspace*{-1cm}
\begin{aligned}
F(r)=&\left(1-\frac{r_-}{r} \right) \left(1-\frac{r_+}{r}\right)
%1-\frac{2\,M}{r}+ \frac{P^2+Q^2}{r^2}
+g_F^2\,\left(\frac{ P\, Q}{ r_{-}
   r_{+}}\right)^2\Biggl\{-\frac{2 r_{-}}{r}+\frac{r_{+}r_{-} }{r^2}-\frac{2r-r_{+}-r_{-}}{r} \log
   \left(1-\frac{r_{-}}{r}\right)\\
   &+\frac{\lambda}{30 r_{-}^4
   r_{+}} \Biggl[\frac{600}{r}-\frac{60 (4 r_{-}+9
   r_{+})}{r^2}-\frac{70 r_{-}
   (r_{-}-3 r_{+})}{r^3}-\frac{30 r_{-}^2
   (r_{-}-2 r_{+})}{r^4}\\
   &-\frac{5 r_{-}^3 (3
   r_{-}-5 r_{+})}{r^5}+\frac{12
   r_{-}^4 r_{+}}{r^6}+\frac{60 \left(10 r^2-9 r
   (r_{-}+r_{+})+8 r_{-} r_{+}\right) \log
   \left(1-\frac{r_{-}}{r}\right)}{r^2
   r_{-}}\Biggr]\Biggr\} \\
%S(r)=& h(r)^{-1}\,F(r)^{-1} \\[10pt]
h(r)=&1+\frac{2\,g_F^2 P^2 Q^2}{r_{-}^2 r_{+}^2} \biggl\{
   \frac{r_{-}}{r-r_{-}}+\log
   \left(1-\frac{r_{-}}{r}\right)\\
   &- \lambda\,\frac{10\,r_{+}}{r_{-}^2\, (r-r_{-})} \biggl[
    1-\frac{r_{-} }{2\,r}-\frac{r_{-}^2}{6\,r^2}-\frac{r_{-}^3 (5 r_{+}-12 r_{-})}{60\,r_{+}\,r^3}-\frac{ r_{-}^4
   }{4\,r^4}-(1-\frac{r}{r_{-}}) \log
   \left(1-\frac{r_{-}}{r}\right)\biggr] \biggr\} 
   %\\[10pt]
%B(r)=&1 \\[10pt]
%D(r)=&0. 
\end{aligned} 
\eeq }

%\beq
%\begin{aligned}
%F(r)=&\,1-\frac{2\,M}{r}+\frac{Q^2+P^2}{r^2}+g^2_F \,\left(\frac{P\,Q}{r_{+}r_{-}}\right)^2 \left[-\frac{2 r_{-}}{r}+\frac{r_{+}r_{-} }{r^2}-\frac{2r-r_{+}-r_{-}}{r}\log\left(1-\frac{r-}{r}\right)\right]\\
%G(r)=&\,F(r)\,\left[
%1+2\,g_F^2\,\left(\frac{P\,Q}{r_{+}r_{-}}\right)^2
 %  \left(\frac{r_{-}}{r-r_{-}}+\log \left(1-\frac{r_{-}}{r}\right)\right)\right]\,.
  % \end{aligned}
%\eeq
Setting $g_F=0$, one finds the standard Reissner-N\"ordstrom solution with magnetic and electric charges. 
 Setting $\lambda=0$ (and leaving  $g_F=0$ switched on) one finds the solution of \cite{Lee:1991jw}. 
The
axion profile backreacts on the metric, with corrections starting only at order $1/r^3$
in an $1/r$ expansion.  Interestingly, the axion profile
 changes the position of the outer horizon of the charged black hole: at leading order in an expansion in the 
 dimensionless couplings $g_F$, $\lambda$, 
 the position of the outer horizon is
 
% \be
 %r_s\,=\,r_+ + g_F^2\,\frac{P^2\,Q^2}{r_{-}^2 r_{+}}\, \left[\log
  % \left(1-\frac{r_{-}}{r_{+}}\right)+\frac{r_{-}}{
  % r_{+}-r_{-}}\right] \,\ge\,r_+
% \ee
 \beq
r_H=r_{+}+g_F^2\, r_S,
\eeq
with
\small{
\beq \label{exp_rs}
\begin{aligned} 
r_S=& \frac{P^2 Q^2}{r_{+}r_{-}^2} \biggl[\log
   \left(1-\frac{r_{-}}{r_{+}}\right)+\frac{r_{-}}{r_{+}-r_{-}}\\
   &\,\,\,\,\,\,\,\,\,\,\,\,\,\,\,\,\,\,\,\,+\lambda\, \frac{r_{-} \left(3
   r_{-}^4+5 r_{-}^3 r_{+}+10 r_{-}^2 r_{+}^2+30 r_{-}
   r_{+}^3-60 r_{+}^4\right)-60 r_{+}^4 (r_{+}-r_{-}) \log
   \left(1-\frac{r_{-}}{r_{+}}\right)}{30 \,r_{-}^3 r_{+}^3\,
   (r_{+}-r_{-})}\biggr]\,.
   \end{aligned}
\eeq  }
 The geometry is regular at the outer horizon,  and we find (in our approximations
 of small $g_F$, $\lambda)$ that the presence of the axion hair  increases the size 
 of the outer horizon.  
Curvature invariants {\it diverge}
 at the position of the inner horizon $r_-$: the axion profile makes singular the inner horizon of the charged
 black hole. This fact was already found in \cite{Campbell:1991rz,Lee:1991jw} for the case with minimal couplings with 
 gravity, and remains true in our set-up with  derivative  couplings to gravity described
 by Lagrangian \eqref{scal_lag}.  
 
  So, as anticipated in Subsection \ref{sec_hairs}, Chern-Simons gauge couplings $\phi F \tilde F$
    provide qualitatively new opportunities to find new black hole solutions with axion hairs, also
    in the case of non-minimal couplings of the axion with gravity. 
 Our analytic  formulas -- valid for small values of the coupling constants --  
  describe how the black hole geometry  
  is affected by the long range axion configurations.  
 
 To conclude this subsection, we qualitatively  compare some aspects of our findings with some of the regular black
 hole solutions in theories of Horndeski. In \cite{Sotiriou:2013qea,Sotiriou:2014pfa}, asymptotically flat, hairy solutions have been found in a system with 
 a linear coupling between the scalar field and the  Gauss-Bonnet  combination (hence
 avoiding the no-hair theorem \cite{Hui:2012qt}, see our Section \ref{sec_hairs}), weighted by a dimensionless   quantity that we call
   $g_{GB}$. 
    Moreover,  \cite{Sotiriou:2014pfa} determines analytical solutions for the system in a perturbative expansion for small  $g_{GB}$,
 finding that the geometry is affected only at next to leading order in such expansion, requiring
 a control of the theory up to next-to-leading level in the  small parameter $g_{GB}$. 
 Instead, in our
 case, differences with respect  to the standard Reissner-N\"ordstrom configuration start already at leading
  order in our parameter expansion. The works 
 \cite{Rinaldi:2012vy,Anabalon:2013oea,Minamitsuji:2013ura}  determine regular black hole solutions
 with non-minimal couplings to gravity like ours, but in order to avoid the no-hair theorem \cite{Hui:2012qt} the
 solutions are not asymptotically flat. A more comprehensive  discussion of black holes in theories
 non-minimally coupled with gravity can be found in the review \cite{Babichev:2017guv}.
 
 \subsection{A Smarr formula for our configurations}
 Since we analytically determined -- at leading order in a 
 perturbative expansion  in the parameters $g_F,\,\lambda$ -- how 
  the axion profile modifies the location of black hole horizon, we can enquire whether
 classical   black hole thermodynamic formulas  remain valid in our system. (See also
 \cite{Jacobson:1993vj} for a general discussion on black hole thermodynamics in theories containing
 non-minimal couplings with gravity  and higher derivative interactions.)
 
   The Smarr formula is  associated
   with the first law of thermodynamics, and relates the black hole entropy  with conserved asymptotic charges.
    It reads, in the context of 
   spherically symmetric geometries \cite{Smarr:1972kt}, 
% 
% Our results are in good agreement with what expected from black hole thermodynamics.
% An (pseudo)scalar profile scaling as $1/r$ is known to influence the first law of black hole thermodynamics,
% with a contribution that reads \cite{Gibbons:1996af}:
\beq
M=\frac{\kappa}{2 \pi}S%+J\,\Omega_H
 +Q\,\Phi_H^E+P\,\Phi_H^M \label{eq:smarr}\,,
\eeq
where $M$ is the ADM mass, %$J$ is the angular momentum, 
 $Q$ and $P$ are the electric and the magnetic charges, $\kappa$ is the surface gravity, $S$ is
 the black hole entropy and $\Phi^{E,\,M}_H$ respectively are the electric and magnetic
 potentials evaluated at the horizon.
%\beq
%\begin{aligned}
%&\kappa:\,\,\,\,\,\,\,\,\,\,\text{surface gravity},\\
%&S:\,\,\,\,\,\,\,\,\, \text{black hole entropy},\\
%&\Omega_H:\,\,\,\,\, \text{black hole event horizon angular velocity},\\
%&\Phi^E_H:\,\,\,\,\, \text{black hole electric potential on the horizon},\\
%&\Phi^M_H:\,\,\,\,\, \text{black hole magnetic potential on the horizon}.\
%\end{aligned}
%\eeq
Using Hawking relation  between the black hole
%
%As it has already been widely pointed out (\cite{Bardeen:73}, \cite{Jacob},\cite{Damour:2004}), 
%The differential version of the Smarr formula (\ref{eq:smarr}) can be regarded as the \emph{First Law} of black holes %thermodynamics, with $\kappa$ playing the same role of the temperature (see e.g \cite{Bardeen:73,Damour:2004}
%\beq
%T_H=\frac{\kappa}{2\pi} \Rightarrow T_H= \frac{\partial E}{\partial S}=\frac{\partial M}{\partial S}.
%\eeq 
%Exploiting this analogy, we can relate the black hole's 
entropy and its event horizon area
\beq
S=\frac{A_H}{4},
\eeq
we can compute  the various quantities appearing in eq \eqref{eq:smarr}. At leading
order in $g_F$, $\lambda$, we find 
\beq
\begin{aligned}
&\kappa =\frac{r_{+}-r_{-}}{2 r_{+}^2}-\frac{g_F^2 P^2 Q^2}{r_{-}^2 r_{+}^4} \biggl[\frac{r_{-}
   (r_{+}-2 r_{-})+(r_{+}-r_{-})^2 \log
   \left(1-\frac{r_{-}}{r_{+}}\right)}{r_{+}-r_{-}}\\
   &\,\,\,\,\,-\lambda\frac{
  \left(r_{-}^5-2 r_{-}^4 r_{+}-5 r_{-}^3 r_{+}^2-20
   r_{-}^2 r_{+}^3+90 r_{-} r_{+}^4-60 r_{+}^5\right)-60\, r_{+}^4\,r_{-} 
 (1- \frac{r_{+}}{r_{-}})^2 \log \left(1-\frac{r_{-}}{r_{+}}\right)}{10\,
   r_{-}^2 \,r_{+}^3\, (r_{+}-r_{-})}\biggr],\\[10pt]
   \end{aligned}
   \eeq
\beq
\begin{aligned}
&A_H=4\pi \left( r_{+}^2+2 \,g_F^2\,r_S\,r_{+}\right),\\[10pt]
%&\Omega_H=0,\\[10pt]
&\Phi^E_H=-A_t(r),\\[10pt]
&\Phi^M_H=\Phi^E_H(P \leftrightarrow Q),
\end{aligned}
\eeq
where $A_t$ is given in eq \eqref{Aprof},  $r_S$ in \eqref{exp_rs},   while the magnetic
potential can be computed using the procedure explained in \cite{Carroll:2004st}, and in our case is
simply related with the electric potential interchanging $P$ with $Q$. 
Substituting these quantities in \eqref{eq:smarr}, we find that the Smarr formula  is satisfied for our black
hole solutions: hence the first law of thermodynamics applies also to dyonic black holes with axion fields
non-minimally coupled with gravity. Notice that formula \eqref{eq:smarr} does not receive `corrections' associated
with   the
pseudoscalar axion field:  a possible reason  is
  that the axion profile does not modify the asymptotic  conserved charges
of our  configuration ($M, P, \,Q$), since it only
 affects metric and gauge fields with higher order corrections in an $1/r$ expansion. In fact,  the
  asymptotic black hole 
  mass and charges do not receive  any  contribution  from the axion, which consequently  
  does not influence the overall energetic balance controlled by the Smarr formula.

\section{Slowly rotating charged configurations}\label{sec_rot}

When turning on spin, besides the gauge, also the gravitational Chern-Simons  coupling
$\phi R \tilde R$ acquires a non-zero value at asymptotic infinity 
 -- see eq \eqref{souR} -- and can
source a non-trivial axion profile. This fact is related with the well-developed
 topic of rotating
black hole solutions in Chern-Simons gravity, see e.g. \cite{Yunes:2009hc,Konno:2009kg,Cambiaso:2010un,Yagi:2012ya,Stein:2014xba,Konno:2014qua,McNees:2015srl,Delsate:2018ome} and the review \cite{Alexander:2009tp}.  

 In this section, we investigate for the first time   how the simultaneous presence 
of gauge and gravitational Chern-Simons couplings affects the geometry of rotating charged black holes
with a long range axion hair. To stress
the role of the gravitational Chern-Simons couplings, we consider only black holes with a magnetic monopole charge (no electric
charge), and
we work in the limit of small rotation, $a\,=\,J/M\ll 1$.  For simplicity, we also turn off the non-minimal couplings
with gravity, $\lambda=0$ in Lagrangian \eqref{scal_lag}. Notice that  this situation  is different from Wald configuration
of a black hole immersed in an external magnetic field \cite{Wald:1974np}, since in our case
it is the black hole itself that is magnetically charged. 

The Ansatz we adopt for the metric is 

  \beq
ds^2\,=\,-F(r)\,dt^2\,+ \frac{dr^2}{F(r)}+ \,r^2\,d\theta^2+ \,r^2\,\sin^2\theta\,d\varphi^2%+D(r)\,dt\,dr
 +2\,a\,w(r)\,r^2\,\sin^2\theta\,dt\,d\varphi,
\eeq
where the quantity $w(r)$ in the off-diagonal component controls the effect of black hole rotation. 
   The gauge vector has the structure
  \beq
A_{\mu}=\Big\{A_t(r,\theta)\,\,0\,,\,0\,,\,A_{\varphi}(r,\theta) \Big\}.
\eeq
  At leading order in the coupling constants $g_F$ and $g_R$, we find the solution

  \beq
\begin{aligned}
F(r)=&\,1-\frac{2\,M}{r}+\frac{P^2}{r^2}\,\\
%S(r)=&\,F(r)^{-1}\\
%B(r)=&\,1\\
%D(r)=&\,0\\
%\end{aligned}
%\eeq
%\beq
%\begin{aligned} 
w(r)=  \,& -\frac{2 M}{r^3}+\frac{P^2}{r^4}-g_F\, g_R\, P^2\, \left(\frac{3}{16 M^2 r^4}+\frac{1}{10 M r^5}+\frac{21}{20 r^6}\right) \\
&+g_R^2 \left[\frac{5}{2 r^6}+\frac{30 M}{7 r^7}+\frac{27 M^2}{4 r^8}-P^2 \left(\frac{3}{16 M^4 r^4}+\frac{9}{40 M^3 r^5}+\frac{3}{10 M^2 r^6}+\frac{41 M}{4 r^9}+\frac{18}{7 M r^7}+\frac{21}{4 r^8}\right)\right] \end{aligned} \label{eq:knn}
\eeq 
and
\beq
\begin{aligned}
A_t(r,\theta)=\frac{a\,P\,\cos\theta}{r^2}\Big[&1+ g_F\,g_R\,\left(-\frac{3}{16 M^2}+\frac{1}{8 M r}+\frac{11}{40 r^2}+\frac{9 M}{20 r^3}\right)\\[7pt]
&-g_R^2 \, \left( \frac{3}{16 M^4}+\frac{3}{16 M^3 r}+\frac{9}{40 M^2 r^2}+\frac{3}{10 M r^3}+\frac{1}{4 r^4}+\frac{9 M}{56 r^5}\right) \Big] ,\\
A_{\varphi}(r,\theta)=-P\,\cos\theta.
\end{aligned}
\eeq
The  associated  profile  for the pseudoscalar axion field has a dipolar structure and results
\small{
\beq
\hspace*{-1cm}
\begin{aligned}
\phi(r,\theta)= &% -g_F\frac{P\,Q }{r_{+}r_{-}}\,\log \left( \frac{r-r_{-}}{r} \right)\\
%&+
\,a \,g_F\frac{
   P^6%\plu P^2 Q^2 \left(P^2-Q^2\right)-Q^6
   }{\left(r_{+}r_{-}\right)^4} \left[2
   (M-r) \log \left(\frac{r-r_{-}
   }{r}\right)\plu \frac{r_{+}r_{-}}{r}-2 \,r_{-}
   \right]\,\cos\theta\\
    & -\frac{8\,a\,g_R}{5} \biggl\{\frac{ 10 M^4-21
   M^2 \left(r_{+}r_{-}\right)\plu 6 \left(r_{+}r_{-}\right)^2}{2\,r
   \left(r_{+}r_{-}\right)^3}-\frac{29 M
   \left(r_{+}r_{-}\right)-15 M^3}{12\,r^2
   \left(r_{+}r_{-}\right)^2}-\frac{18 \left(r_{+}r_{-}\right)-5
   M^2}{12\,r^3 \left(r_{+}r_{-}\right)}\\
  &- \frac{5 M}{4\,r^4}\plu \frac{
   \left(r_{+}r_{-}\right)}{r^5}+\frac{ 10 M^4-21 M^2
   \left(r_{+}r_{-}\right)\plu 6 \left(r_{+}r_{-}\right)^2
  }{\left(r_{+}r_{-}\right)^4} \left[(M-r) \log \left(\frac{r-r_{-} }{r}\right)-r_{-} \right]
   \biggr\}
\,\cos\theta.   \label{eq:HAIRKN}
 \end{aligned}
 \eeq}
 {Interestingly,} it includes contributions from both the gauge and gravitational Chern-Simons terms, 
 weighted by the (small) rotation parameter $a$.
 
At our level of approximations, the position of the black hole horizon {and consequently  the black hole thermodynamics} 
are not modified with respect to the  Reissner-N\"ordstrom magnetised solution.
On the other hand, the geometry and the gauge field receive non-trivial corrections in the metric
coefficient $w(r)$ and in the gauge component $A_t(r,\theta)$, which acquire  new `mixed' contributions proportional
to $g_F \,g_R$ due to the simultaneous presence
of gauge and gravitational Chern-Simons terms. As far as 
we are aware, this is the first time that such mixed
contributions have been found in this context,   showing there is an interplay between
gauge and gravitational Chern-Simons terms for determining the black
hole geometry with axion hairs.  Thanks
to this contribution, our solution generalises to a rotating,
charged
setting similar configurations  discussed in the works \cite{Campbell:1990ai}, \cite{Boskovic:2018lkj}.

\section{Phenomenological considerations}\label{sec_app}

\subsection{Black hole axion hairs and light polarisation}\label{sec_pheno_1}

In the previous Sections, we have seen  that charged black holes can 
develop long range axion hairs, thanks to the gauge
and gravitational Chern-Simons couplings contained in action \eqref{eq:ACTION}. The Abelian charges carried by
the black holes are not necessarily  electromagnetic, and they can be associated with some additional gauge group motivated
by string theory constructions,
or dark forces associated
with dark matter interactions. 

 Even if the black hole configurations are  charged under extra Abelian
 gauge groups that {\it do not correspond} to  the electromagnetic $U(1)$ symmetry, 
  we make the hypothesis
 that  axions additionally  couple also with electromagnetic photons, through a coupling 
 \be \label{coupfot}
 g_{\rm EM}\,\phi\,\mathcal{F}_{\mu\nu}\, \tilde{\mathcal{F}}^{\mu\nu}\,,
 \ee
with $\mathcal{F}_{\mu\nu}$ being the electromagnetic $U(1)$ field
strength and  $g_{\rm EM}$ the coupling constant
controlling the axion-photon interaction. The previous formula  denotes a dimension five operator, that we
weight with appropriate powers of the Planck mass (as in the previous Sections) and regard
$g_{\rm EM}$ as a dimensionless coupling. 
 (In case
the black holes are electromagnetically charged, then
$g_{\rm EM}\,=\,g_{F}$ with $g_{F}$  the coupling
%coincides with the $g_{F}$ 
we analysed in the previous sections.)
%only if the $U(1)$ gauge group charging the black
%hole is the electromagnetic $U(1)$ -- otherwise
%it is a new independent parameter. 

A possible way to observationally detect a pseudoscalar black hole hair   is to measure polarised  bending of light
 travelling in proximity of the black hole.
 %\footnote{Surprisingly, as far as we are aware, there are not many known astrophysical events which could produce right and left circularly polarised waves.} 
    In fact, the parity violating  coupling \eqref{coupfot} bends by a different amount left-handed and right-handed photon trajectories crossing the pseudoscalar profile, by a magnitude depending also on the values 
 of $g_{EM}$, $g_F$, $g_R$: so,  quantifying this effect could help in probing 
 the magnitude of the Chern-Simons couplings characterising the theory. 
 
  This effect has been  studied in detail in the recent work \cite{Plascencia:2017kca} for the case of  photons  travelling through the  cloud of light axions   depositing on a rotating black hole ergosphere. % the phenomenon causing black hole superradiance. %
 Instead, here we study the same effect for light travelling through
 the long range axion profile studied in the previous Sections. We use the same approach and methods of 
 \cite{Plascencia:2017kca}, hence we refer to the reader to \cite{Plascencia:2017kca} for more details on the derivations. We study two cases, spherically symmetric and slowly rotating black holes, which can probe different sets of parameters.

\begin{itemize} 

\item {\bf Spherically symmetric black hole configurations}

 \medskip
 
 \noindent
We study the bending of light passing in proximity of  spherically symmetric, dyonic black hole configurations with long range
axion hairs,  discussed in Section \ref{sec_spher}. At large distances from the black hole horizon, the black
hole geometry can be described by a Schwarzschild configuration (the corrections associated with the
axion hair backreaction starts only at order $1/r^3$ in a large $r$ expansion), with a long range axion hair profile, 
that scales with radius as in eq \eqref{scal_asympt}:
\be
\Phi_{\rm dyon}(r)\,=\,\frac{g_F\,P\,Q}{r_+}\,\frac{1}{r}+{\cal O}\left(\frac{1}{r^2} \right)\,, \label{eq:dyonhair}
\ee
with $r_+$ the position of the external horizon in a dyonic Reissner-N\"ordstrom configuration (see eq \eqref{old_hor}). 
We assume that  light passes  sufficiently far from the black hole, so  that a Schwarzschild approximation
for the geometry is valid and for simplicity we assume that  the photon trajectory lies in the equatorial plane of the black hole. Calling $E_\gamma$
and $L$ the conserved photon energy and angular momentum, and $r$ and $\varphi$
respectively the radial and angular polar coordinates controlling the photon path on the black
hole equatorial plane (the black hole being
located at $r=0$),  we find the orbit equation
\beq
\frac{d \varphi_\pm}{dr}\,=\,\frac{1}{r^2}\left[\frac{\pm g_{\rm EM}}{2L} \left(1-\frac{2 M}{r}\right){ \Phi'_{\rm dyon}(r )}+\sqrt{\frac{E_{\gamma}^2}{ L^2}+\frac{2
   M-r}{r^3}+ \frac{g_{\rm EM}^2}{4L^2} \frac{(2M-r)^2}{r^2}{ \Phi'^{\,2}_{\rm dyon}(r )}}\right]^{-1}, \label{eq:obbi}
\eeq
where $\pm$ refer to the two photon polarisations, while $ \Phi_{\rm dyon}$ is the axion profile given in 
eq \eqref{eq:dyonhair}. The contributions proportional to $g_{\rm EM}$ are associated with the axion-photon
interactions: {remarkably, in the 
presence of a long range axion profile  they can `distinguish' among the two different photon polarisations.}  
The closest radial approach $r_0$ of the light orbit to the black hole configuration {is given by the ratio of the energy and the angular momentum:}
%\cite{Plascencia:2017kca}%
\beq
\frac{E_{\gamma}}{L}=\frac{1}{r_0}\sqrt{\frac{r_0-2 M}{r_0}}\,. \label{eq:closa}
\eeq
Starting from these formulas, we can straightforwardly determine the photon deflection angle as travelling from spatial
infinity to the distance to closest approach to the black hole. We are  especially interested on the total
 difference $\Delta  \varphi_\pm$  in the angular deflection between right and left polarisation. 
Calling $x_0\,=\,r_0/M$, we find that this quantity is described by a simple formula
\beq
|\Delta  \varphi^{\rm charged}_\pm |=\frac{(2x_0-3)}{x_0^{5/2}\sqrt{x_0-2}  }\,\,\,\frac{g_F\,g_{\rm EM}}{6E_{\gamma}\,M^2}\,\,\,\frac{P\,Q}{r_{+}}. \label{eq:phANGd}
\eeq
A measurement of this quantity can then probe the electromagnetic coupling of the axion to photons, and
the amount of charge of a black hole configuration. 

 \medskip
 
 \noindent
\item {\bf Slowly rotating  black hole configurations}

 \medskip
 
 \noindent
 We now study polarised light deflection around the slowly rotating configurations  examined
 in Section \ref{sec_rot}. We would like to focus
 on the effects of purely gravitational
 Chern-Simons couplings, hence we switch off the black hole magnetic
 and electric charges.  At large distances from the black hole horizon, the geometry (at leading order in a $1/r$
 expansion) is still well described by a Schwarzschild solution, and the axionic profile corresponds to a dipolar
 configuration given by
 \beq
\Phi_{\rm rot}(r,\theta)\,=\,\frac{5}{8}\frac{a}{M} \frac{ g_R}{r^2} \cos\theta + {\cal O} \left( \frac{1}{r^2}\right)\,\,.
\eeq
 Also in this case we study photon trajectories as done above, but -- since the axion configuration is a dipole -- we consider trajectories
 lying not on the equatorial plane, but on the more convenient plane $\theta\,=\,\pi/6$. Proceeding
 as we explained in the spherically symmetric case, we find that
  the angular deflection of polarised photons
   is governed by the following equation
   \beq
\frac{d \varphi_{\pm}}{dr}=\frac{2}{r_0^2}\left[\pm\alpha 
   \left(1-\frac{2 M}{r}\right) \sqrt{\frac{1-\frac{2
   M}{r_0}}{r_0^2}}+\sqrt{-\frac{1-\frac{2 M}{r}}{r^2}+ \frac{1-\frac{2 M}{r_0}
   }{r_0^2}\left(1+ \frac{\alpha ^2 (r-2
   M)^2}{r^2}\right)}\right]^{-1}. \label{eq:linkerr}
\eeq
where $r_0$ is the  closest approach distance of the photon to the black hole -- see eq. \eqref{eq:closa} -- and
\beq
\alpha=\frac{g_{\rm EM}}{2\,E_{\gamma}}\frac{\partial \Phi_{\rm rot}(r,\pi/6)}{\partial r}. \label{eq:DEFKN}\,
\eeq
We learn that also in this case the axion profile  bends
differently the two photon polarisations. Integrating over all the trajectory, the difference
$\Delta  \varphi_\pm$  in the angular deflection between right and left polarisations now reads
\beq
|\Delta  \varphi^{\rm rot}_\pm|\,
=
\,\frac{\sqrt{3}}{16}\,\frac{(5x_0-8)}{x_0^{7/2}\sqrt{x_0-2}} \frac{a \, g_R\,g_{\rm EM}}{E_{\gamma}\,M^4}. \label{eq:def}
\eeq
Interestingly, eq \eqref{eq:def}   can probe a different set of parameters
in this case, with respect to spherically  symmetric charged configurations (compare with
eq \eqref{eq:phANGd}). 

\end{itemize}
To summarise, we find compact formulas, eqs \eqref{eq:phANGd} and \eqref{eq:def}, for  
the difference in light bending for the two  polarisations of a photon travelling within
a region containing black hole  axion hairs, as described in Sections \ref{sec_spher} and \ref{sec_rot}.  
These formulas can probe different parameters: while in the spherically symmetric
case the polarised light bending can probe gauge couplings with Chern-Simons terms,
in the slowly rotating case we can also probe gravitational Chern-Simons interactions -- 
 a coupling that as far as we are aware is not probed by direct axion experiments.   
%Formulas \eqref{} depend on a variety of  parameters, and if these effects
%are  observed they can allow to  measure the coefficients of Chern-Simons couplings
%$g_{\rm EM}$, $g_F$, $g_R$ we discussed in this work.  
 Future astronomical observations based on  radio astronomy can achieve
an angular resolution of order $\Delta \phi\sim 10^{-4}$ \cite{Plascencia:2017kca}: it would be
interesting to quantitatively estimate at what extent these precise
measurements can test the existence and properties of black hole axionic hairs,
and their couplings with electromagnetic photons.  We plan to return
on these topics soon.

\subsection{ISCO}

As a second application of our findings, we  compute 
 the ISCO (Innermost Stable Circular Orbit) of the dyonic black hole configuration 
 with axion hairs as
   discussed in Section \ref{sec_spher}, including the effects of non-minimal couplings 
 of axions with gravity, Lagrangian \eqref{scal_lag}.
  The properties of ISCOs can be interesting for a phenomenological characterization
  of the geometrical features of the system under investigation. Since they depend on
  the properties of the configuration relatively near the black hole horizon, they can
  be sensitive to the effects of Chern-Simons couplings, or of the derivative 
  non-minimal couplings with gravity.  
  By defining
 \beq
f(r)=\log[F(r)],\,
\eeq
 with $F(r)$ being the coefficient of the time-time metric component, see eq \eqref{met_sol1}, 
  the ISCO radius is given by the root   \cite{Barausse:2011pu,Sotiriou:2014pfa} of  the following equation
\beq
3\,f'(r)-r\,f'(r)^2+r\,f''(r)^2=0\,.
\eeq
We can solve the previous condition analitically by expanding for small values of  magnetic and electric charges $P$ and
$Q$.
% We introduce  the following expansion parameter
%\beq
%\epsilon^2=P^2+Q^2.
%\eeq
%Explicitly, the ISCO equation reads
%\beq
%\small{
%\begin{aligned}
%   &-\frac{2 M (6 M-r)}{r (r-2 M)^2}-\frac{2 M (6 M-7 r)}{r^2 (r-2
%   M)^3}\,\epsilon ^2 +\frac{ \left(4 M^2+2 M r-8 r^2\right)}{r^3
%   (r-2 M)^4}\,\epsilon ^4+\frac{2 
%   \left(2 M^2-9 M r+8 r^2\right)}{r^4 (r-2
%   M)^5}\,\epsilon ^6\\
%&+g_F^2 P^2 Q^2 \left[\frac{-16 M^2+10 M r+3 r^2}{4 M r^3
%   \left(144 M^3-298 M^2 r+205 M r^2-48 r^3\right)}{30 M^2 r^7
%   (2 M-r)^3}-\frac{24 M^4+208 M^3 r-234 M^2 r^2+20 M r^3+3
%   r^4}{24 M^3 r^4 (r-2 M)^4}\right)\right]=0
  % \end{aligned}}
%\eeq
%
 %Expanding for small values of $\epsilon$, 
 We push our  perturbative
 expansion up to  order  $\mathcal{O}({\rm charge}^6)$ for  catching the effects of the Chern-Simons
 coupling $g_F$, and non-minimal couplings with gravity $\lambda$.
In these approximations, the ISCO radius
 results %{\bf \color{red} here we should explain how to pass from the previous to the next formula}
 \bea
r_{ISCO}&=&6M-\frac{\left( P^2+Q^2\right)}{2M}-\frac{19\,\left( P^2+Q^2\right)^2}{72M^3}-\frac{5
\left( P^2+Q^2\right)^3}{48M^5} 
\nonumber
\\
&+&g_F^2 P^2\,Q^2\,
\left[ \frac{19}{144} +\left( P^2+Q^2\right)\,\frac{(49410M^2-193\lambda)}{466560M^7}\right].
\eea
The first line contains the General Relativity expression for the ISCO radius: the Schwarzschild result ($6M$) and
the first corrections in small values of the charges associated
with a dyonic Reissner-N\"ordstrom configuration. The second line contains instead the contributions 
associated with the presence of black hole axion hair,  the Chern-Simons terms, and non-minimal
coupling with gravity. For small values of the charges, these contributions start only     at order
 $\mathcal{O}({\rm charge}^4)$, hence it can be difficult  to use the properties
  of ISCO to reveal  the existence of axion hairs. 
To conclude, we  report the result for the angular frequency associated with the ISCO trajectory
\small{
\beq
\begin{aligned}
\omega_{ISCO}=&+\frac{1}{6 \sqrt{6} M}+\frac{7\,
   \left( P^2+Q^2\right) ^2}{144 \sqrt{6} M^3}+\frac{49 \,\left( P^2+Q^2\right) ^4}{2304 \sqrt{6} M^5}+\frac{5489\,
  \left( P^2+Q^2\right) ^6}{497664 \sqrt{6} M^7}\\
&+g_F^2 P^2 \,Q^2 \left[-\frac{1}{216 \sqrt{6} M^5}+\left( P^2 + Q^2\right) ^2 \left( \,\frac{11\,\lambda}{699840 \sqrt{6}
   M^9}-\frac{47}{7776 \sqrt{6} M^7}\right)\right].
   \end{aligned}
   \eeq}
Again, the first term correspond to the Schwarzschild result, while the contributions from the scalar field start to appear at order $\mathcal{O}({\rm charge}^4)$.

\section{Conclusions}

We studied spherically symmetric and slowly
rotating charged black hole configurations with long range axion hairs. We focussed on
 Einstein-Maxwell theories equipped with an axion Lagrangian
that preserves a shift symmetry for the axion field. 
Gauge and gravitational Chern-Simons 
couplings,  eq \eqref{eq_intro}, are essential 
for evading no-hair theorems, and lead to long range  axion profiles.  We extended known black
hole 
solutions to cases    where additional  derivative couplings of axion to curvature are switched on,
and to situations
in which both gauge and gravitational Chern-Simons couplings are
present simultaneously. 
 In all cases, we determined analytical solutions at leading order in the coupling constants involved, determining  how
 the axion profile   backreacts  on the metric and the gauge field.  The metric remains regular outside
 the outer horizon, and the position
 of such horizon is increased with respect to  dyonic solutions in Einstein-Maxwell gravity with the same
 asymptotic conserved charges. Moreover, the solution for the electric potential is  modified with respect
 to the standard case due to the effect of  the gauge Chern-Simons couplings. To make contact with phenomenology,
 we studied two possible consequences of our findings. First, we studied how   axion hairs can induced
  polarised bending of photons  travelling within the range of the axion profile around a black hole. Then, we
  investigated the properties of ISCOs around the spherically symmetric
  configurations we analysed. It would be interesting, in the future, to  study the stability of the systems considered
  under small perturbations of the fields involved, and investigate possibly parity breaking effects -- induced by the Chern-Simons
  couplings -- in the dynamics of  fluctuations around these geometries.

\subsection*{Acknowledgments}
It is a pleasure to thank Ivonne Zavala for discussions. We are partially supported by STFC grant ST/P00055X/1.

\begin{appendix}
\section{An exact solution for large  values of $\lambda$}\label{app_a}
 In the main text, we presented a solution for a spherically symmetric dyonic black hole configuration
  at leading order in a perturbative expansion in the coupling constant $\lambda$, characterising
  derivative couplings of the axion to the curvature, introducted eq \eqref{scal_lag}.  Solutions for arbitrary
  values of $\lambda$ can be found in general, but their expressions are very cumbersome. On the other
  hand, for special values of this parameter, their expression simplifies. For example,
  with the particular choice
\beq
\lambda=-\frac{r_{-}^3}{2\,r_{+}} \label{eq:lambdasp}
\eeq
the solution for the axion  configuration
 is relatively simple, and reads
% of motion simplifies a lot, and the solution for the axion field reads 
%\small{
\beq
\phi(r)=\frac{g_F\,P\, Q}{8
   r_{-} r_{+}} \left[ \frac{2\,r_{-}}{r-r_{-}}+\log \left(r^2+r_{-}^2\right) +\log
   (r+r_{-})-3 \log (r-r_{-})+2 \tan
   ^{-1}\left(\frac{r_{-}}{r}\right)\right].
\eeq
Notice that the choice \eqref{eq:lambdasp} allows for tuning large values of $\lambda$, by choosing $r_+$ and
$r_-$ appropriately. 
%Due to the non trivial profile of the axion field, it is interesting to study the backreaction of the field with the metric. \\
 The metric line element is
\beq
ds^2=-F(r)\,dt^2+\frac{dr^2}{h(r)F(r)}+r^2\,d\theta^2+r^2\sin^2\theta\,d\varphi^2.
\eeq
Choosing  $\lambda$ as in eq \eqref{eq:lambdasp},
 at leading order in the coupling $g_F$ we find
 % have
\vspace{5mm}
\small{\beq
\hspace*{-0.8cm}
\begin{aligned}
F(r)=&\left(1-\frac{r_{+}}{r} \right) \left( 1-\frac{r_{-}}{r}\right)+\frac{g_F^2\,P^2\,Q^2}{8r_{-}^2\,r_{+}^2}\biggl[ \frac{ 2r_{-} \left( 4 r_{+}-3r \right)}{r^2} +\frac{(r r_{-}+r r_{+}-2 r_{-} r_{+})}{r^2}\log
   \left(r^2+r_{-}^2\right)\\
   &+\frac{\left(r^2-2 r
   r_{+}-r_{-} r_{+}\right)}{r^2} \log (r+r_{-})-\frac{\left(r^2+2 r r_{-}-5
   r_{-} r_{+}\right)}{r^2} \log (r-r_{-})+  \frac{ 2(2 r-r_{-}-3 r_{+})}{r} \tan
   ^{-1}\left(\frac{r_{-}}{r}\right) \biggr],\\[10pt]
h(r)=&1-\frac{g_F^2\,P^2\,Q^2}{2 r_{+}^2} \biggl[\, \frac{3 r^8 r_{+}+r^7 r_{-} r_{+}+r^6 r_{-}^2
   r_{+}+r^5 r_{-}^3 (8 r_{-}+r_{+})-13 r^4
   r_{-}^4 r_{+}-r^3 r_{-}^5 r_{+}-r^2
   r_{-}^6 r_{+}-r r_{-}^7 r_{+}+2 r_{-}^8
   r_{+}}{2 r_{-} r_{+} (r-r_{-})^3
   (r+r_{-})^2 \left(r^2+r_{-}^2\right)^2}\\
   &\,+\frac{2\tan ^{-1}\left(\frac{r_{-}}{r}\right)+\tanh
   ^{-1}\left(\frac{r_{-}}{r}\right)}{ r_{-}^2}\biggr].
   \end{aligned}
\eeq}
As in the small coupling approximation $\lambda \ll 1$ (see section \ref{sec_hairs}), the Reissner-N\"ordstrom case is recovered setting $g_F=0$, while metric corrections due to axion's backreaction start only at order $1/r^3$ in an $1/r$ expansion.  
At leading order in $g_F$ expansion, the position of the outer horizon becomes
\beq
r_H=r_{+}+g_F^2\,r_S
\eeq
with
\vspace{5mm}
\beq
\begin{aligned}
r_S=\frac{P^2 Q^2}{8 r_{-}^2 r_{+}
  } \Biggl[ &-\frac{2 r_{-}}{ (r_{+}-r_{-})}
   -\log
   \left(r_{-}^2+r_{+}^2\right)+\frac{(r_{+}-3 r_{-})}{ (r_{+}-r_{-})}
   \log (r_{+}-r_{-})+\frac{(r_{-}+r_{+})}{ (r_{+}-r_{-})} \log
   (r_{-}+r_{+})\\
   &+ \frac{2(r_{-}+r_{+})}{ (r_{+}-r_{-})} \tan
   ^{-1}\left(\frac{r_{-}}{r_{+}}\right)\Biggr].
   \end{aligned}
\eeq
With the particular choice (\ref{eq:lambdasp}), compared with the Reissner-N\"ordstrom case the axion's backreaction increases the size of the outer horizon. It would be interesting to understand if there is any choice of the coupling $\lambda$ which could lead to a reduction of the size of the outer horizon, but we leave it to future works.
The geometry is regular on the outer horizon $r_H$ and everywhere outside it, while there is a singularity located on the inner horizon $r_{-}$, which cannot be removed by any choice of the coupling parameter $\lambda$.

Moreover, the axion's backreaction also modifies the gauge potential
\beq
A_{\mu}=\{ A_t(r),\,0,\,0,\,A_{\varphi}(\theta) \},
\eeq
and at leading order in a $g_F$ expansion we find
\vspace{5mm}
\small{
\beq
\hspace{-1cm}
\begin{aligned}
A_t(r)=&-\frac{Q}{r}+ g_F^2 \,Q\Biggl\{\frac{P^2 Q^2}{8 r_{-}^2
   r_{+}^2} \Biggl[\frac{r^5 (3 r_{-} + r_{+ })+ r^4 r_{-}
   (r_{+ }-r_{-})+ r^3 r_{-}^2
   (r_{+ }-r_{-})+ r^2 r_{-}^3
   (r_{+ }-r_{-})+ 4 r r_{-}^5-8 r_{-}^5
   r_{+ }}{4 r  r_{+ } (r-r_{-})^2
   (r+ r_{-}) \left(r^2+ r_{-}^2\right)}\\
   &-\frac{(r_{-}-5 r_{+ })
   \log \left(r^2+ r_{-}^2\right)}{4 r_{-}
   r_{+ }}+ \frac{(5 r
   r_{-}-13 r r_{+ }+ 4 r_{-} r_{+ }) \log
   (r-r_{-})}{8 r r_{-} r_{+ }}-\frac{(r
   r_{-}+ 7 r r_{+ }+ 4 r_{-} r_{+ }) \log
   (r + r_{-})}{8 r r_{-} r_{+}}\\
   &+ \frac{\tan^{-1}\left(\frac{r}{r_{-}}\right)}{ r_{-}}\Biggr]
   +\frac{P^2}{4\,r_{-}^2r_{+}} \Biggl[  \frac{(r_{-}-r) \log
   \left(r^2+ r_{-}^2\right)}{2r }+ \frac{(r-3 r_{-}) \log (r-r_{-})}{2 r}+ \frac{(r+ r_{-}) \log
   (r+ r_{-})}{2 r }\\
   &-\frac{r \tan
   ^{-1}\left(\frac{r}{r_{-}}\right)+ r_{-}}{ r} + \frac{P^2}{r_{+}} \frac{\tan
   ^{-1}\left(\frac{r_{-}}{r}\right)}{ r 
   } \Biggr]\Biggr\} ,\\[10pt]
A_{\varphi}(\theta)=&-P\,\cos\theta.
\end{aligned}
\eeq}
As for the perturbative solution in the main text, the magnetic potential is the same as in Einstein-Maxwell
theory, while the electric potential is modified by the presence of the axion field.
\end{appendix}

\addcontentsline{toc}{section}{References}
\bibliographystyle{utphys}

\bibliography{refsBHs}

\end{document}